\documentstyle[aps,manuscript]{revtex}
\newcommand{\beq}{\begin{equation}}
\newcommand{\eeq}{\end{equation}}
\def\beqa{\begin{eqnarray}}
\def\eeqa{\end{eqnarray}}

\def\lap{\lower.5ex\hbox{$\; \buildrel < \over \sim \;$}}
\def\gap{\lower.5ex\hbox{$\; \buildrel > \over \sim \;$}}

\begin{document}
\title{Gamma ray bursts from superconducting cosmic strings}

\author{V. Berezinsky$^{1)}$, B. Hnatyk$^{1),2)}$ and A. Vilenkin$^{3)}$}

\address{$^{1)}$INFN, Laboratori Nazionali del Gran Sasso, I-67010 Assergi
(AQ), Italy \\
$^{2)}$Institute for Applied Problems in Mechanics and Mathematics,
NASU, Naukova 3b, Lviv-53, 290053, Ukraine\\
$^{3)}$Physics Department, Tufts University, Medford, MA 02155, USA}

\maketitle
\begin{abstract}
Cusps of superconducting strings can serve as GRB engines. A powerful
beamed pulse of electromagnetic radiation from a cusp produces a jet of
accelerated particles,
whose propagation is terminated by the shock responsible for
GRB. A single free parameter,
the string scale of symmetry breaking $\eta \sim  10^{14}~GeV$,
together with reasonable assumptions about the magnitude of
cosmic magnetic fields and the fraction of volume that
they occupy,
explains the GRB rate, duration
and fluence, as well as the observed ranges of these
quantities. The wiggles on the string can drive the short-time structures
of GRB.
This model predicts that GRBs are accompanied by strong bursts of
gravitational radiation which should be detectable by LIGO, VIRGO
and LISA detectors. Another prediction is the diffuse X- and gamma-ray
radiation at 8~MeV - 100~GeV with a spectrum and flux comparable
to the observed.  The weakness of the model is the prediction of
too low rate of GRBs from galaxies, as compared with observations.
This suggests that either the capture rate of
string loops by galaxies is underestimated in our model, or that GRBs from
cusps are
responsible for only a subset of the observed GRBs not associated
with galaxies.

\end{abstract}

\section{Introduction}
Existing models of gamma ray bursts (GRBs) face the problem of
explaining the
tremendous energy released by the central engine \cite{Pir98}.
In the case of isotropic emission, the total energy output should be as
high as $1.4\times 10^{54}$ ergs, in case of GRB 990123 with redshift
$z=1.6$. Strongly beamed emission is needed for all known
engine models, such as mergers and hypernovae, but such extreme
beaming is difficult to arrange (see recent discussion by
Blandford \cite{Bland} and Rees \cite {Rees84}).
In this
paper we show that emission of pulsed electromagnetic radiation
from cusps of superconducting cosmic
strings naturally solves this problem and explains
the observational GRB data using only one engine parameter \cite{BHV}.

Cosmic strings are linear defects that could be formed at a symmetry
breaking phase transition in the early universe \cite{Book}.  Strings
predicted in most grand unified models respond to external
electromagnetic fields as thin superconducting wires \cite{Witten}.
As they move through cosmic magnetic fields, such strings develop
electric currents.
Oscillating loops of
superconducting string emit short bursts of highly beamed
electromagnetic radiation through small string segments, centered at
peculiar points on a string, cusps, where velocity reaches speed of light.
\cite{VV,SPG}.

The idea that GRBs could be produced at cusps of
superconducting strings was first
suggested by Babul, Paczynski and Spergel \cite{BPS} (BPS) and further
explored by Paczynski \cite{Paczynski}.
They assumed that the bursts
originate at very high redshifts ($z\sim 100 - 1000$),  with GRB
photons produced either directly or in electromagnetic cascades
developing due to interaction with the microwave background.
This model requires the existence of a strong primordial magnetic
field to generate the string currents.

As it stands, the BPS model does not agree
with observations.  The observed GRB redshifts are in the range
$z\lesssim 3$, and the observed duration of the bursts ($10^{-2}s
\lesssim \tau\lesssim 10^3 s$) is significantly longer than that
predicted by the model.  On the theoretical side, our understanding of
cosmic string evolution  and of the GRB generation in relativistic jets
have considerably evolved since the BPS papers
were written.  Our goal in this paper is to revive the BPS idea taking
stock of these recent advances.

As in the BPS model we shall use the cusp of a
superconducting string
as the central engine in GRB. It provides the tremendous engine
energy naturally beamed.  Our main observation is that putting
superconducting cusps in a different enviroment, the magnetized
plasma at a relatively small redshift $z$, results in a different
mechanism of
gamma radiation, which leads to a good agreement with GRB
observational data.

GRB radiation in our model arises  as follows.
Low-frequency electromagnetic radiation
from a cusp
loses its energy by accelerating particles of the plasma
to very large Lorentz factors.
Like the initial electromagnetic pulse, the particles are beamed
and give rise to
a hydrodynamical flow in the surrounding gas, terminated by a shock,
as in the standard fireball theory of GRB \cite{MeRee92}
(for a review see
\cite{Pir98}).

The string symmetry breaking
scale $\eta$ will be the only string parameter used in our
calculations. With reasonable assumptions about the magnitude of
cosmic magnetic fields and the fraction of volume in the Universe that
they occupy,
this parameter is sufficient to account for all main
GRB observational quantities:
the duration $\tau_{GRB}$, the rate of events $\dot N_{GRB}$, and the
fluence $S$.

We begin in the next Section with a brief review of cosmic string
properties and evolution, with an emphasis on the physics of cusps and
on the generation and dissipation of electric current in
superconducting strings.  (The discussion of the latter topic in the
existing literature is often oversimplified and sometimes incorrect,
so we review it in more detail than we otherwise would.)  The GRB
characteristics in our model are calculated in Section III, and the
hydrodynamical aspects of the model are discussed in Section IV.  In
Section V we discuss the diffuse X-ray and $\gamma$-ray backgrounds
predicted by the model, as well as other observational predictions,
which include GRB repeaters, bursts of gravitational radiation, and
ultrahigh-energy particles.

\section{String overview}

\subsection{String properties and evolution}

Here we briefly review some aspects of cosmic string properties and
evolution, which are relevant for
the discussion below (for a detailed review and references see \cite{Book}).

Strings are characterized by the energy scale of symmetry breaking
$\eta$, which is
given by the expectation value of the corresponding Higgs field,
$\langle\phi\rangle=\eta$.  The mass per unit length of string is
given by
\beq
\mu\sim\eta^2.
\label{mu}
\eeq
An important dimensionless parameter characterizing the gravitational
interactions of strings is
\beq
G\mu\sim (\eta/m_P)^2,
\label{Gmu}
\eeq
where $G$ is Newton's constant and $m_P$ is the Planck mass.  In many
models this is the only relevant string parameter.

Numerical simulations of cosmic string evolution indicate that strings
evolve in a self-similar manner \cite{BB,AS,AT}.
A horizon-size volume at any time $t$
contains a few long strings stretching across the volume and a large
number of small closed loops.  The typical distance between long
strings and their characteristic curvature radius are both $\sim t$,
but, in addition, the strings have small-scale wiggles of wavelength
down to
\beq
l\sim\alpha t,
\label{l}
\eeq
with $\alpha\ll 1$.  The typical length of loops being chopped off the
long strings is comparable to the scale of the smallest wiggles
(\ref{l}).

The loops oscillate periodically and lose their energy, mostly by
gravitational radiation.  For a loop of invariant length $l$
\cite{invlength}, the
oscillation period is $T_l=l/2$ and the lifetime is
$\tau_l\sim l/k_g G\mu$, where $k_g\sim 50$ is a numerical coefficient.

The exact value of the parameter $\alpha$ in (\ref{l}) is not known.
Numerical simulations give only an upper bound, $\alpha\lesssim
10^{-3}$, while the analysis of gravitational radiation
backreaction indicates that $\alpha\gtrsim k_g G\mu$.
We shall assume, following \cite{BB}, that $\alpha$ is determined by the
gravitational backreaction, so that
\beq
\alpha\sim k_g G\mu.
\label{alpha}
\eeq
Note that in this case the loops decay within about one Hubble time of
their formation.  Then, most of the loops at time $t$ have length
$l\sim\alpha t$, and their number density is given by
\beq
n_l(t)\sim \alpha^{-1}t^{-3}.
\label{n}
\eeq

Analysis of string equations of motion reveals that oscillating loops
tend to form cusps, where for a brief period of time the string
reaches the speed very close to the speed of light.  Near a cusp, the
string gets contracted by a large factor, its rest energy being turned
into kinetic energy.  For a string segment of invariant length $\delta
l\ll l$, the maximum contraction factor is $\sim l/\delta l$, resulting
in a Lorentz factor
\beq
\gamma\sim l/\delta l.
\eeq

To avoid confusion, we note that cusps were originally
defined \cite{Turok84} as points of infinite contraction, where the string
momentarily reaches the speed of light. Strictly speaking, such cusps can
be formed only on idealized infinitely thin strings.
For realistic strings, the
cusp development is truncated either by the annihilation of overlapping
string segments at the tip of the cusp\cite{Branden87,jjkdo98.0,jjkdo98.1}
or for superconducting strings, by the back reaction of charge
carriers or of the electromagnetic radiation. However,
unless the string current is very large, so that the energy
of the charge carriers is comparable to that of the string itself, the
truncation occurs at a very large Lorentz factor and the string exhibits
cusp-like behavior. Below we shall use the word ``cusps'' to refer to such
ultra-relativistic string segments.

Cusps typically form a few times during an oscillation period, but
it is possible to construct (somewhat contrived) loop configurations
exhibiting no cusps.  Apart from various backreaction effects, the
motion of loops is strictly periodic, and thus cusps reappear at
nearly the same locations on the string in each oscillation period.

Another peculiar feature that one can expect to find on string loops
is a kink \cite{Garfinkle}. It is characterized
by a sharp bent, where the string direction changes discontinuously.
Two oppositely
moving kinks are produced on a loop at the moment when the loop is
disconnected from a long string.  The kinks then run around the loop
at the speed of light.

\subsection{String superconductivity}

As first shown by Witten \cite{Witten}, strings predicted in a wide
class of elementary particle models behave as superconducting wires.
If some fermions acquire their mass as a result of the same symmetry
breaking that is responsible for the string formation, then these
fermions are massive outside the string but are massless inside.
If in addition some of these fermions are electrically charged, then
the strings have massless charge carriers which travel along the
string at the speed of light.  The fermion mass outside the string is
$m=\lambda\eta$, where $\lambda$ is the Yukawa coupling of the fermion
to the Higgs field of the string.  Yukawa couplings in particle
physics models are often very small, so it is not unusual to have
$m\ll \eta$.
String superconductivity can also be
bosonic, with charge carriers being either spin-zero bosons or
spin-one gauge particles.  Here, we shall consider
only fermionic superconductivity.

An electric field $E$ applied along a superconducting string generates
an electric current.  The Fermi momentum of the charge carriers grows
with time as ${\dot p}_F=eE$, where $e$ is the elementary charge,
and the number of fermions per unit length,
$n=p_F/2\pi$, also grows, ${\dot n}\sim eE$.  The resulting current,
$J\sim en$, grows at the rate
\beq
dJ/dt \sim e^2E.
\label{dJdt}
\eeq

A superconducting loop oscillating in a magnetic field $B$
acts as an {\it ac} generator and
develops an {\it ac} current of amplitude
\beq
J_0\sim e^2Bl.
\label{J}
\eeq
This loop current is not homogeneous; it changes direction along the
string and is more accurately described as current-charge
oscillations.  Some portions of the loop develop charge densities
$\sim J_0$.
For typical values used in the calculations below,
$B=1\cdot 10^{-7}$~G and $l=\alpha t_0\sim 30$~pc,
with $\alpha = 1\cdot 10^{-8}$ and $t_0 \sim 10^{10}~$yrs the present
age of the universe, one obtains
$J_0\sim 2\cdot 10^5~{\rm GeV}$.

The local value of the string current can be greatly enhanced in the
vicinity of cusps.  The portion of the string that attains a Lorentz
factor $\gamma$ is contracted by a factor $\sim 1/\gamma$.  The charge
carrier density, and thus the current, are enhanced by the same
factor, so the current becomes (in the local rest frame of the string)
\beq
J_\gamma\sim\gamma J_0.
\label{Jgamma}
\eeq

The growth of electric current at the cusp  is
terminated at a critical value $J_{max}$ when the energy of charge
carriers becomes comparable to that of the string itself, $(J/e)^2\sim
\mu$.  This gives $J_{max}$ and $\gamma_{max}$ as \cite{BOV}
\beq
J_{max}\sim e\eta, ~~~~\gamma_{max}\sim (e\eta/J_0).
\label{Jmax}
\eeq
Alternatively, the cusp development can be terminated by small-scale
wiggles on the string \cite{Carter}.  If the wiggles contribute a
fraction $\epsilon\ll 1$ to the total energy of the string, then the
maximum Lorentz factor is less than (\ref{Jmax}), and is given by
\beq
\gamma_{max}\sim\epsilon^{-1/2}.
\eeq
The actual value of $\gamma_{max}$ is not
important for most of the following discussion.

In realistic models, the strings have several fermion species as
charge carriers.  It can be shown that fermions of a given species can
move only in a certain direction along the string.  Thus, the charge
carrier species can be divided into left-movers and right-movers.  If,
for example, the applied electric field is directed to the right,
it produces positively charged right-movers and negatively charged
left-movers (and vice versa for the opposite direction of $E$).
The left-movers and right-movers usually differ
by flavour, lepton number, or some other conserved or weakly violated
quantum number.

In the absence of an external electromagnetic field, the current in an
oscillating loop decays due to various dissipation mechanisms.  These
include: scattering of left- and right-movers
\cite{Barr,Shafi}, electromagnetic back-reaction
\cite{Aryal,SPS}, and plasma effects \cite{Thompson}.

Charge carrier loss due to scattering of left- and right-movers is
highly model-dependent.  If the scattering is mediated by superheavy
gauge bosons of mass $M_X\sim 10^{15}$~GeV,
then the characteristic scattering time is \cite{Barr}
\beq
\tau_{sc}\sim 3\cdot 10^4\left({J\over{10^5~{\rm GeV}}}\right)^{-5}
~{\rm yrs}.
\eeq
For $J<10^4$~GeV this time is greater than the age of the universe,
but $\tau_{sc}$ decreases rapidly with the growth of the current and
becomes comparable to the typical oscillation period of loops ($T_l
\sim 100$~yrs for $l\sim 30$~pc) for $J\sim 3\cdot 10^5$~GeV.
In near-cusp regions, where $J\gg 10^5$~GeV, charge carrier
scattering becomes very efficient.

We note, however, that this current loss mechanism has an important
limitation.  The densities of left- and right-moving charge carriers
are typically not equal, and even if scattering were 100\% efficient,
it would stop after eliminating the minority charge carriers, leaving
the string with a chiral current (that is, with a current consisting
of only left- or right-movers).  This is what we expect to happen in
the vicinity of cusps.

The electromagnetic back-reaction typically damps the loop current on
a timescale $\tau_{em}\sim l/e^2\sim 100 l$, which is much shorter
than the loop's lifetime.  It tends to damp the spatial component of
the current, with the total charge of the loop remaining the same, so
in the absence of other effects the end result would be left- and
right-moving currents of the same magnitude and the same
charge.\footnote{Spergel {\it et. al.} \cite{SPS} argued that the {\it
dc} component of the current cannot be changed by the e-m
back-reaction.  However, their Eq. (11) which they quote in support of
this statement is in fact an expression of charge conservation.}
Combined with scattering of charge carriers, this mechanism can
dissipate loop charges and currents, even in the chiral case.
Moreover, the string charge is almost completely screened by a vacuum
condensate \cite{Nascimento},
so the string is effectively neutral even if the
scattering rate is low and there is some residual charge.
It should be noted that the physics of electromagnetic back-reaction
can be significantly modified by plasma effects, which are presently
not well understood.  Thompson \cite{Thompson} has argued that current
damping becomes more efficient in the presence of plasma.

Another mechanism that can dissipate a large chiral current operates
when a loop oscillates in an external magnetic field.  The emf induced
in the loop oscillates with the same period.  Suppose for definiteness
that the loop initially has a chiral current $J_i$ consisting of
positively charged right-movers.  When the emf is directed oppositely
to this current, the magnitude of the right-moving current is reduced
by $\sim J_0$ and a positively charged left-moving current of
magnitude $\sim J_0$ is generated, with $J_0$ from Eq. (\ref{J}).
Left- and right-movers can now scatter off the string, and if
$\tau_{sc} < T_l$, the chiral component of the current will be reduced
by $\sim J_0$.  The initial current will then be dissipated in $\sim
J_i/J_0$ oscillations.

The effect of all these dissipation mechanisms is to damp the loop's
charge and current on a timescale
\beq
\tau_d\sim (1 - 100)l.
\label{taud}
\eeq
This means in particular that the loop quickly
forgets any initial charge or current that it inherits when it is
chopped off the long string network.  The magnitude of the current in
a loop is determined mainly by the local magnitude of the cosmic
magnetic field, as in Eq. (\ref{J}).

We note finally that Eq. (\ref{J}) for the current is
modified when the loop has an appreciable center-of-mass velocity
$v$ \cite{SPG}.  In this case, the amplitude of current-charge
oscillations grows
linearly with time, until the growth is hampered by the damping
processes.  The resulting amplitude is
\beq
J_0\sim e^2 Bv\tau_d.
\eeq
Loops can have high center-of-mass velocities, $v\sim 1$, but in view
of the uncertainty in the damping time (\ref{taud}) we shall use the
estimate (\ref{J}) for the current.

\section{GRB engine}

There are three types of
sites in the universe where magneic fields can induce
large electric currents in the strings. They are compact structures
(galaxies and clusters of galaxies), voids, and  walls (filaments
and sheets) of the large-scale structures. The total rate of GRBs is
dominated by the walls, and further on we shall concentrate on
these structures only.

Magnetic fields in our scenario are assumed to be generated in young
galaxies during the bright phase of their evolution \cite{ParPee} and then
dispersed by galactic winds in the intergalactic space. Then
at present the fields are concentrated in the filaments and
sheets of the large-scale structure \cite{Bier,param}.

Assuming that magnetic fields were generated at some
$z\sim z_B$ (galaxy formation epoch) and then remained frozen in the
extragalactic plasma, we obtain
\beq
B(z)=B_0(1+z)^2,
\label{B}
\eeq
where the characteristic field strength at the present time $B_0$
can be estimated from the equipartition condition as
$B_0 \sim 10^{-7}$~G \cite{Bier}.

With sheets of
characteristic size $L\sim (20-50)h^{-1}$ Mpc and thickness $D\sim
5h^{-1}$ Mpc, we can estimate the fraction of the space occupied  by
the walls with magnetized plasma  as  $f_B\sim D/L\sim 0.1$.
For numerical estimates below we shall use $z_B\sim 4$.

We shall now estimate the physical quantities characterizing GRBs
powered by cusps of superconducting strings.  In what follows we
assume that the universe is spatially
flat, is dominated by non-relativistic matter, and has age
$t_0= 0.87\cdot 10^{10}$~yr, which corresponds to
dimensionless Hubble constant $h=0.75$.

\subsection{GRB rate and fluence}

Due to the large current, the cusp produces a
powerful pulse of electromagnetic radiation. The total energy of the pulse
is given by \cite{VV,SPG} $ {\cal E}_{em}^{tot} \sim 2 k_{em}J_0 J_{max}l$,
where $l \sim \alpha t$ is the length of the loop, and the coefficient
$k_{em} \sim 10$
is taken from numerical calculations \cite{VV}.
This radiation is emitted within a very narrow cone of
openening  angle $\theta_{min} \sim 1/\gamma_{max}$.
The angular distribution of radiated
energy at larger angles is given by \cite{VV}
\beq
d{\cal E}_{em}/d\Omega \sim k_{em}J_0^2 l /\theta^3.
\label{Emax}
\eeq

For a GRB originating at redshift $z$ and seen at angle $\theta$ with
respect to the string velocity at the cusp, we have from
Eqs.(\ref{J})-(\ref{B})
\beq
d{\cal E}_{em}/d\Omega
\sim k_{em}e^4\alpha^3 t_0^3 B_0^2 (1+z)^{-1/2}\theta^{-3},
\label{Eobs}
\eeq
The Lorentz factor of the
relevant string segment near the cusp is $\gamma \sim 1/\theta$.
The duration of the cusp event as seen by a distant observer is \cite{BPS}
\beq
\tau_c\sim (1+z)(\alpha t/2)\gamma^{-3}
\sim (\alpha t_0/2) (1+z)^{-1/2}\theta^{3}.
\label{tau}
\eeq
One can expect that the observed duration of GRB is
$\tau_{GRB}\sim\tau_c$.  This expectation will be justified by the
hydrodynamical analysis in Section IV.

The fluence, defined as the total energy
per unit area of the detector, is \cite{Paczynski}
\beq
S\sim (1+z)(d{\cal E}_{em}/d\Omega) d_L^{-2}(z),
\label{S}
\eeq
where $d_L(z)=3t_0(1+z)^{1/2}[(1+z)^{1/2}-1]$ is the luminosity
distance.

The rate of GRBs originating at cusps in the redshift interval $dz$
and seen at an angle $\theta$ in the interval $d\theta$ is given by
\beq
d{\dot N_{GRB}}\sim f_B\cdot {1\over{2}}\theta d\theta
(1+z)^{-1}\nu(z) dV(z).
\label{dN}
\eeq
Here, $\nu(t)\sim n_l(t)/T_l\sim 2\alpha^{-2}t^{-4}$ is the number of
cusp events per unit spacetime volume, $T_l\sim \alpha
t/2$
is the oscillation period of a loop,
$dV=54\pi t_0^3[(1+z)^{1/2}-1]^2(1+z)^{-11/2} dz$ is the proper volume between
redshifts $z$ and $z+dz$,
and we have used the relation $dt_0=(1+z)dt$.

Since different cusp events originate at different redshifts and are
seen at different angles, our model automatically gives a distribution
of durations and fluences of GRBs.  The angle $\theta$ is related to
the Lorentz factor of the relevant portion of the string as
$\theta\sim 1/\gamma$, and
from Eqs.(\ref{Eobs}),(\ref{S}) we have
\beq
\gamma (z;S) \sim
\gamma_0 \alpha^{-1}_{-8}S_{-8}^{1/3}B_{-7}^{-2/3}
[(\sqrt{1+z}-1)^2\sqrt{1+z}]^{1/3}.
\label{gamma}
\eeq
Here, $\gamma_0 \approx 190,~~\alpha_{-8}=\alpha/10^{-8}$, and the
fluence $S$ and the magnetic field $B_0$ are expressed
as $S=S_{-8}\cdot 10^{-8}~erg/cm^2$ and $B_0=B_{-7}\cdot 10^{-7}~G$.

Very large values of $\gamma\sim\gamma_{max}$, which correspond
(for a given redshift) to largest fluences, may not be seen at all because
the radiation is emitted into a too narrow solid angle and the observed
rates of these events are too small.
The minimum value $\gamma(z;S_{min})$
is determined by the smallest fluence that is observed,
{\em e.g.} for GRBs at $z\gtrsim 1$
with  $S_{min}\sim 3\cdot 10^{-8}~{\rm erg/cm^2}$,
~~ $\gamma_{min}\approx 170$.
Another lower limit on $\gamma$, which dominates at small $z$, follows
from the condition of compactness \cite{Pir98} and is given by $\gamma
\gtrsim 100$ (see Section IV).

The total rate of GRBs with fluence larger than $S$ is obtained by
integrating Eq.(\ref{dN})
over $\theta$ from
$\gamma_{max}^{-1}(z)$ to $\gamma^{-1}(z;S)$ and over $z$ from $0$
to $\min[z_m;z_B]$, with $z_m$ from $\gamma_{max}(z_m)=\gamma(z_m;S)$.
For relatively small fluences,
$S_{-8}<S_c=0.03(\gamma_{max}(0)\alpha_{-8}/\gamma_0)^3B_{-7}^2$,
$z_B<z_m$
and we obtain
\begin{eqnarray}
{\dot N_{GRB}(>S)} &\sim& \frac{f_B}{2\alpha^2 t_0^4 }
\int_0^{z_B}dV(z)(1+z)^5\gamma^{-2}(z;S) \nonumber\\
&\sim& 3\cdot 10^2 S_{-8}^{-2/3}B_{-7}^{4/3}~yr^{-1}.
\label{rate}
\end{eqnarray}
Remarkably, this rate in our model does not depend on any string
parameters and is
determined (for a given value of $S$) almost entirely by the
magnetic field $B_0$.  It agrees with the observed rate for
$B_{-7}\sim 1$ (formally, the observed rate
$\dot{N} \sim 300 \mbox{yr}^{-1}$ at $S>1\cdot 10^{-7}\mbox{erg/cm}^2$
gives $B_{-7}=3.2$).
The predicted slope $\dot N_{GRB}(>S) \propto S^{-2/3}$
is also in a reasonable agreement with the observed one
${\dot N}_{obs}(>S)\propto S^{-0.55}$ at relatively small fluences
\cite{Bie99}.

For large fluences $S_{-8} >S_c$,
integration of Eq.(\ref{dN}) gives
$\dot N_{GRB}(>S) \propto S^{-3/2}$. Observationally, the transition
to this regime occurs at $S_{-8}\sim 10^2-10^3$.  This can be
accounted for if the cusp
development is terminated by small-scale wiggles with fractional
energy in the wiggles
$\epsilon\sim 10^{-7}\alpha_{-8}^2 B_{-7}^{4/3}$.
Alternatively, if $\gamma_{max}$ is
determined by the back-reaction of the charge carriers,
Eq.(\ref{Jmax}), then the regime
(\ref{rate}) holds for larger $S_{-8}$, and observed steepening of the
distribution at large $S$ can be due to the reduced efficiency of BATSE
to detection of
bursts with large $\gamma$. Indeed, large $\gamma$ results in
a large Lorentz factor $\gamma_{CD}$ of the emitting region (see
Section IV), and
at $\gamma_{CD} \gtrsim 10^3$ photons start to escape from the
BATSE range.

\subsection{GRB duration}

The duration of GRBs originating at redshift $z$ and having fluence
$S$ is readily calculated from Eqs.(\ref{tau}) and (\ref{gamma}) as
\beq
\tau_{GRB} \approx 200 \frac{\alpha_{-8}^4 B_{-7}^2}{S_{-8}}
(1+z)^{-1}(\sqrt{1+z}-1)^{-2}~s
\label{duration}
\eeq

 From Eqs.(\ref{dN}) and (\ref{tau}) we find the rate of GRBs with
durations in the interval $d\tau$ and redshifts in the
interval $dz$,
\beq
d{\dot N}\sim 10^2\alpha^{-2}t_0^{-1}\left({\tau\over{\alpha
t_0}}\right)^{2/3}(\sqrt{1+z}-1)^2 (1+z)^{-1/6} dz{d\tau\over{\tau}}.
\label{dNtau}
\eeq
The distribution of GRB durations is found by integrating this over $z$.
The integration is restricted by $z<z_B$ and $S> S_{min}\sim 3\cdot
10^{-8} ~{\rm erg/cm^2}$.  The latter
condition can be expressed as $z < {\tilde z}(\tau)$, where ${\tilde
z}(\tau)$ is the solution of Eq. (\ref{duration}) for $z$ with $S\sim
S_{min}$.


The distribution changes its form at the
characteristic value $\tau_*$ defined by ${\tilde
z}(\tau_*) =z_B$.  With $z_B=4$, Eq. (\ref{duration}) gives
\beq
\tau_*\sim 8.7\alpha_{-8}^4 B_{-7}^2 ~{\rm s}.
\label{tau*}
\eeq
For $\tau<\tau_*$ we have $d\dot{N} \propto \tau^{2/3}d\tau/\tau$, and for
$\tau \gg \tau_*$, ~ $d\dot{N} \propto \tau^{-5/6}d\tau/\tau$.
We thus see that the distribution is peaked at
$\tau\sim\tau_*$.

The largest value of $\tau$ in our model is obtained from Eq.(\ref{tau})
with $\theta\sim\theta_{max}\sim 10^{-2}$,
$\tau_{GRB}^{max}\sim 10^3\alpha_{-8} ~{\rm s}$.
There is no sharp lower cutoff, but very small values of $\tau$ will
not be observed due to the low rate of events.
With the rate $\sim 10^2$~yr$^{-1}$ near the peak of the distribution,
the rate of events with $\tau\sim 10^{-4}\tau_*$ is about $0.1$~yr$^{-1}$.

A lower
bound on $\tau$ is also set by the detector resolution ($\sim
10^{-2}$ s for BATSE).  Hence, we have
$\tau_{min}\sim {\rm max}\{ 10^{-4}\tau_*,10^{-2}~{\rm s}\}$.

The observed distribution of GRB durations extends from $\sim
10^{-2}$ s to $\sim 10^3$ s.  The distribution is bimodal, with
peaks at 0.5 s and 15 s \cite{Yu}, and there are some observational
indications that short and long GRBs may have different origin.
Our model is probably
better suited to describe the short GRB population (see Section V).
With $\tau_*\sim 0.5$~s and $B_{-7}\sim 3$, Eq.(\ref{tau*}) gives
$\alpha_{-8}\sim 0.3$.  This
corresponds to the string symmetry breaking scale $\eta\sim 1\cdot
10^{14} ~{\rm GeV}$.  The range of
GRB durations is then given by $\tau_{GRB}^{min}\sim 10^{-2}~{\rm s}$,
$\tau_{GRB}^{max}\sim 10^3~{\rm s}$.

It should be noted that the validity of our simple one-parameter model
does  not extend beyond rough order-of-magnitude estimates (see
Section VI).  In particular, it is not expected
to give the correct duration distribution ${\dot N}(\tau)$, and
identifying the peaks of the theoretical and observed distributions
may therefore exceed the accuracy of the model.  A more conservative
approach is to require that the characteristic duration $\tau_*$ lies
within the observed range of GRB durations.  This gives $0.2 \lesssim
\alpha_{-8} \lesssim 3$.

\section{Acceleration and hydrodynamics}

A beam of low-frequency e-m radiation propagating in plasma
produces a beam of accelerated particles.

The characteristic
frequency of e-m radiation in a pulse produced by a cusp segment
with Lorentz factor $\gamma$ is
\beq
\omega_{em}\sim \frac{4\pi}{\alpha t_0}\gamma^3(1+z)^{3/2}=
4.6(\gamma/10^3)^3(1+z)^{3/2}\alpha_{-8}^{-1}~\mbox{s}^{-1}.
\label{om-em}
\eeq
The plasma frequency in the intergalactic gas of density
$n=n_{-5}10^{-5}$~cm$^{-3}$,
\beq
\omega_{pl}=1.8\cdot 10^2 n_{-5}^{1/2}~\mbox{s}^{-1} ,
\label{om-pl}
\eeq
is higher than $\omega_{em}$ when
$\gamma \lesssim 3.4\cdot 10^3n_{-5}^{1/6}\alpha_{-8}^{1/3}(1+z)^{-1/2}$.
Therefore, the low-frequency radiation from the cusp cannot propagate
in plasma. In fact, the energy density of e-m beam is much
larger than that of the
plasma, and the beam would push the plasma away even in the case
$\omega_{em} > \omega_{pl}$. This process occurs due to the acceleration
of plasma particles.

Let us consider the propagation of a charged test particle in a
strong, low-frequency e-m wave.
For a time interval much shorter than the period of the wave,
$t\ll1/\omega_{em}$, the e-m field of the wave can be approximated by
static, orthogonal electric and magnetic fields of equal magnitude.
Solution of the equations of motion (see e.g. \cite{LL}) shows
that both positive and negative charges are accelerated mainly in the
direction of wave propagation, ${\bf n}= ({\bf E \times B})/EB$,
with their Lorentz factor increasing with time as
\beq
\gamma_b(t)=\left( \frac{3}{\sqrt{2}}\frac{eB}{m}t\right)^{2/3},
\label{gammab}
\eeq
where $m$ is the particle's mass.
The synchrotron energy loss of an accelerated particle is small, because
when it moves in the direction of wave propagation, ${\bf n}$, the
electric force, $e{\bf E}$, and the magnetic force $e{\bf v\times B}$
almost exactly compensate each other:
$e({\bf E}+v{\bf n\times B}) \approx 0$. This regime of
acceleration is practically the same as in the Gunn-Ostriker
mechanism \cite{Gunn}.

For an e-m wave in vacuum,
a test particle would be
accelerated at $t \sim 1/\omega_{em}$ up to a very large Lorentz
factor. But the maximum Lorentz factor of the {\em beam}  is saturated
at the value $\gamma_b$, when the energy of the
beam reaches the energy of the original e-m pulse:
$N_b m \gamma_b \sim {\cal E}_{em}$. This results in the Lorentz factor
of the beam
\beq
\gamma_b \sim 4\cdot 10^2 B_{-7}^{2}n_{-5}^{-1}(1+z)^4(\gamma/100)^6.
\label{gamma_b}
\eeq

Let us now turn to the hydrodynamical phenomena in which the gamma radiation
of the burst is actually generated.
The beam of accelerated particles pushes the
gas with the frozen magnetic field ahead of it, producing an external
shock in surrounding plasma  and  a reverse  shock
in the beam material, as in the case of ``ordinary'' fireball
(for a review see \cite{Pir98}). The difference is that the beam propagates
with a very large Lorentz factor $\gamma_b \gg \gamma$,
where $\gamma$ is the Lorentz factor of the cusp (the precise value of
$\gamma_b$ is not important for our discussion).
Another difference is that the beam
propagates in a very low-density gas. The beam can be regarded as a
narrow shell of relativistic particles of width $\Delta \sim l/2\gamma^3$
in the observer's frame.

The gamma radiation of the burst
is produced as synchrotron radiation of electrons accelerated by
external and reverse shocks. Naively, the duration of synchrotron
radiation, i.e. $\tau_{GRB}$, is determined by the thickness of the
shell as $\tau_{GRB} \sim \Delta$.  This is confirmed by a more
detailed analysis, as follows.
The reverse shock in our case is ultrarelativistic
\cite{KobPirSar98,Pir98}.
The  neccessary
condition for that, $\rho_b/\rho < \gamma_b^2$, is satisfied with a
wide margin
(here $\rho_b$ is the baryon density in the beam and $\rho$ is the
density of unperturbed gas).  In this case,
the shock dynamics and the GRB duration are determined by two
hydrodynamical parameters \cite{Pir98}.
They are the thickness of the shell $\Delta$
and the Sedov length, defined as the distance travelled by the shell
when the mass of the snow-ploughed
gas becomes comparable to the initial energy of the beam. The latter is
given by $l_{Sed}\sim({\cal E}_{iso}/\rho)^{1/3}$.

The reverse shock enters the shell and, as it propagates there, it strongly
decelerates the shell. The synchrotron radiation occurs mainly in
the shocked regions of the shell and of the external plasma.
The surface separating these two regions,
the contact discontinuity (CD) surface,  propagates with
the same velocity as the shocked plasma, where the GRB radiation is
produced.

The synchrotron radiation ceases when the reverse shock reaches the
inner boundary of the shell. This occurs at a distance
$R_{\Delta} \sim l_{Sed}^{3/4}\Delta^{1/4}$
when the Lorentz
factor of the CD surface is
\beq
\gamma_{CD}\sim (l_{Sed}/\Delta)^{3/8}
\sim 0.1 B_{-7}^{1/4}n_{-5}^{-1/8}(1+z)^{1/2}\gamma^{3/2}.
\label{gammaCD}
\eeq
Note that these
values do not depend on  the Lorentz factor of the beam $\gamma_b$
and are determined by the cusp Lorentz factor $\gamma$.
The size of the synchrotron emitting region is of the order $R_{\Delta}$,
and the Lorentz factor of this region is equal to $\gamma_{CD}$.
The compactness condition \cite{Pir98} requires $\gamma_{CD}\gtrsim
10^2$, and Eq. (\ref{gammaCD}) yields $\gamma\gtrsim 10^2$ which we
used earlier in Section III.
The duration of GRB is given by
\beq
\tau_{GRB}\sim {R_\Delta}/{2\gamma_{CD}^2}\sim {l}/{2\gamma^3},
\label{taugrb}
\eeq
{\em i.e.} it is equal to the duration of the cusp event given by
Eq.(\ref{tau}).
The energy that goes into
synchrotron radiation is comparable to the energy of the
electromagnetic pulse.

\section{Predictions and problems}

In this section we shall consider a number of predictions of our
model.  Some of these predictions pose potential problems.\\*[2mm]
\noindent
{\em (i) Short-time structure of GRBs}.

Most of GRBs exhibit a complex  short-time structure. These
variations must be a property of inner engine \cite{Pir98,Pir97}.
In the cusp model they can be naturally produced by wiggles.
Wiggles are amplified in near-cusp regions and, acting like
minicusps, produce a sequence of successive fireballs.
A quantitative analysis of this effect would require a detailed study of
the gravitational back-reaction, which controls the amplitude of the
wiggles.\\*[2mm]
{\em (ii) Repeaters}.

Cusps reappear on a loop with a period of loop oscillation, producing
nearly identical GRBs.
In our model, where all loops have the same
length $l=\alpha t$ at a given cosmological epoch $t$, the recurrence time,
$T_l \sim (1+z)\alpha t/2 \sim 50\alpha_{-8}(1+z)^{-1/2}~yr$, is too long
to be observed by BATSE and other detectors.
In a more realistic models, some fraction of loops would have
lengths smaller than $\alpha t$ and thus shorter recurrence periods.
This fraction is model-dependent. Moreover, GRBs from repeaters with
$l< \alpha t$ must be weak and have short durations.

The GRB fluence from a loop of length $l$ produced by a string segment with
Lorentz factor $\gamma$ can be readily calculated as
\beq
S\sim\frac{1}{9}k_{em}e^4l^3\gamma^3\frac{B_0^2}{t_0^2}
\frac{(1+z)^4}{[(1+z)^{1/2}-1]^2}
\eeq
 After a change of variables from $\gamma$ and $l$ to $\tau_{GRB}$ and
the reccurence
period $T_{rec}=l(1+z)/2$, we obtain
\beq
\tau_{GRB}\sim k_{em}e^4\frac{B_0^2}{t_0^2}
\frac{(1+z)}{[(1+z)^{1/2}-1]^2}\frac{T_{rec}^4}{S}.
\label{repeat}
\eeq
A search for repeaters with $T_{rec} \leq 5$~yr requires, according
Eq.(\ref{repeat}), low fluences $S \lesssim 10^{-7}$ erg/cm$^2$ and
short durations $\tau_{GRB} \lesssim 40$~ms. The BATSE efficiency is low for
such events \cite{Nemir98} and the repeating burst could
easily have been lost.  The
total number of GRBs shorter than 40 ms in BATSE 3B catalogue is less
than 5 \cite{Cline99}.\\*[2mm]
{\em (iii) Host galaxies}

The discovery of GRB aferglows revealed an association of long-duration
GRBs with galaxies (see \cite{Pa00} for a review).
19 GRBs with long durations are found to be undoubtedly hosted by normal
galaxies \cite{Bl00}. For 10 of them redshifts are found to be
typically 1 - 3.
For many bright bursts, which are most probably at small
distances, no host galaxies have been found. For example, for 16
bright bursts observed by the Interplanetary Network with small error
boxes, no galaxies are found with magnitudes from 20 to 24.  This
suggests that some of the GRBs are not hosted by galaxies.

In our
model, the fraction of loops captured by galaxes is expected to be
small, due to the high velocities of the loops.
The most straightforward way to reconcile the model with observations
is to assume that cusps are responsible only for a subset of the
observed GRBs not associated with galaxies. Such a subset could include
the short-duration GRBs, for which no host galaxies have
yet been detected.
With the choice of parameters
$B_{-7}\sim 3$, $\alpha_{-8}\sim 0.3$, as in Section III, the
distribution of GRB durations is peaked at $\tau_{GRB}\sim 0.5 ~{\rm
s}$.  At the same time, the tail of the distribution extends all the
way to $\tau_{GRB}\sim 10^3 ~{\rm s}$, and thus the model can account
for {\it some} of the long GRBs as well.
This particular possibility meets  another problem, since
short GRBs do not show deviation from Euclidean distribution. However,
it is often suggested that GRBs comprise a few subclasses, and
the existence of a no-host subclass
remains plausible.


An alternative possibility should be also mentined.
In string evolution models with $\alpha\gg k_g G\mu$, the
lifetime of the loops is $\tau_l\gg t$, so the loops will be slowed
down by the expansion of the universe and a substantial fraction of
them can be captured in galaxies.  \\*[2mm]
{\em (iv) Bursts of gravitational radiation}

Our model predicts that GRBs should be accompanied by
strong bursts of gravitational radiation (see also \cite{DaVi}).  The
angular distribution of
the gravitational wave energy around the direction of the cusp is
\cite{VV84} $d{\cal E}_g/d\Omega\sim G\mu^2 l/\theta$, and the
dimensionless amplitude of a burst of duration $\tau$
originating at redshift $z$ can be estimated as
\beq
h\sim k_g^{-1/2}\alpha^{5/3}(\tau/t_0)^{1/3}(1+z)^{-1/3}z^{-1},
\eeq
or $h\sim 10^{-21}\alpha_{-8}^{5/3}z^{-1}(\tau/1s)^{1/3}$ for $z\lesssim
1$.  Here, we have used the relation $F_g\sim h^2/G\tau^2 \sim (1+z)
(d{\cal E}_g/d\Omega)/ d_L^2 \tau$ for the gravitational wave flux and
Eq.(\ref{tau}) for the burst duration $\tau$.
These gravitational wave bursts are much stronger
than expected from more conventional
sources and should be detectable by the planned LIGO, VIRGO and LISA
detectors.  It has been shown in \cite{DaVi} that gravitational
wave bursts from strings are linearly polarized and have a
characteristic waveform $h(t)\propto t^{1/3}$.\\*[2mm]
{\em (v) X- and $\gamma$- ray diffuse radiation}

Tremendous energy [see Eq.(\ref{Emax})] released in a narrow angle
$\theta \sim 1/\gamma_{max}$ is not seen in GRBs because of the
smallness of this angle. The beam of particles accelerated by e-m
radiation in this narrow cone has a very large Lorentz factor,
and the emitted photons have energies in excess of 1~TeV. These
photons are absorbed in collisions with infrared (IR)
or microwave photons, collectively
denoted as  $\gamma_t$: $\gamma+\gamma_t\to e^++e^-$. Electrons and positrons
start e-m cascades on microwave photons ($\gamma_{bb}$) due to Inverse
Compton scattering ($e+\gamma_{bb} \to e+\gamma$) and pair production
($\gamma+\gamma_{bb} \to e^++e^-$). As they degrade in energy, cascade
electrons are effectively deflected in the extragalactic magnetic field,
and the produced diffuse gamma radiation is isotropic. The spectrum of
remaining cascade photons was calculated analytically in \cite{cascade}
(for recent Monte Carlo simulation see \cite{cascadeMC}). The analytic
spectrum is described in terms of three parameters: $\epsilon_{\gamma}$,
$\epsilon_X$ and $\omega_{\gamma}$.

$\epsilon_{\gamma}$ is the minimum energy of absorption, {\em i.e.}
the smallest
energy of a photon absorbed on IR radiation
($\epsilon_{\gamma} \sim m_e^2/\varepsilon_{IR}$, with the exact value
dependent on the spectrum of IR radiation). $\epsilon_X$ is the
energy of an IC photon produced by an electron of energy
$\epsilon_e=\epsilon_{\gamma}/2$, {\em i.e.} by an electron born
by a photon of energy $\epsilon_{\gamma}$. $\omega_{\gamma}$ is
the energy density of cascade radiation.

The space density of cascade photons, $n_{\gamma}(E)$, is given by
\cite{cascade}
\beq
n_{\gamma}(E)=\left\{ \begin{array}{ll}
K(E/\epsilon_X)^{-3/2} & \mbox{if}~~ E\leq \epsilon_X\\
K(E/\epsilon_X)^{-2}   & \mbox{if}~~ \epsilon_X\leq
E \leq \epsilon_{\gamma}\\
0                        & \mbox{if}~~ E>\epsilon_{\gamma}
\end{array}
\right.
\eeq
where $K$ is a normalization constant which can be expressed in terms of
$\omega_{cas}$ as
\beq
K=\frac{\omega_{cas}}{\epsilon_X^2(2+\ln\epsilon_{\gamma}/\epsilon_X)}
\eeq
The cascade energy density $\omega_{cas}$ can be calculated as the
total energy release in e-m radiation of cusps integrated over redshits
from 0 up to $z_B=4$. This gives
\beq
\omega_{cas}=5f_B k_{em}e^3\eta B_0/t_0
\eeq
Assuming $\epsilon_{\gamma}\approx 100$~GeV due to absorption on IR radiation,
we obtain $\epsilon_X \approx 8.1$~MeV. Then the predicted spectrum
in the energy range $8~\mbox{MeV}\leq E\leq 100$~GeV is
\beq
I_{theor}(E)\sim 2.5\cdot 10^{-10}\alpha_{-8}^{1/2}B_{-7}
\left( \frac{E}{10^3\mbox{MeV}}\right) ^{-2}~cm^{-2}s^{-1}sr^{-1}MeV^{-1}.
\label{Ith}
\eeq
This is to be compared with the EGRET flux \cite{EGRET} for the energy range
$5~\mbox{MeV}\leq E\leq 100~\mbox{GeV}$,
\beq
I_{obs}(E)=1.38\cdot10^{-9}\left (\frac{E}{10^3~\mbox{MeV}}
\right)^{-2.1 \pm 0.03}~ cm^{-2}s^{-1}sr^{-1}MeV^{-1}.
\label{Iexp}
\eeq
With $\alpha_{-8} \sim 0.3$ and $B_{-7}\sim 3$
the predicted flux differs from the observed one
by a factor of 3. This can be regarded as agreement for an order of
magnitude estimate of our simple model.\\*[2mm]

{\em (vi) Ultra High Energy Cosmic Rays}

GRBs have been suggested as possible sources of the
observed ultrahigh-energy cosmic rays (UHECR) \cite{Vie,Wax}.
This idea encounters two difficulties. First, if GRBs are
distributed uniformly in the universe, UHECR have
a classical  Greisen-Zatsepin-Kuzmin (GZK)
cutoff, absent in the observations. Second, the acceleration by an
ultrarelativistic shock is possible only in the one-loop regime ({\em i.e.}
due to a single reflection from the shock) \cite{GalAch}.
For a standard GRB with a
Lorentz factor $\gamma_{sh} \sim 300$ it results in the maximum energy
$E_{max} \sim \gamma_{sh}^2 m_p \sim 10^{14}~eV$,  far too low for UHECR.

Our model can resolve both of these difficulties, assuming that
$\gamma_{max}$ is determined by the current backreaction, Eq.(\ref{Jmax}).

If the magnetic field in the Local Supercluster (LS) is considerably
stronger than outside, then the cusps in LS
are more powerful and the GZK cutoff is less pronounced.

Cusp segments with large Lorentz factors produce
hydrodinamical flows with large Lorentz factors, {\em e.g.},
$\gamma \sim 2\cdot 10^4$ corresponds to $\gamma_{CD} \sim 3\cdot
10^5$ and
$E_{max} \sim \gamma_{CD}^2m_p \sim 1\cdot 10^{20}~eV$. Protons with
such energies are deflected in the magnetic field of LS and can be
observed, while protons with much higher energies caused by near-cusp
segments with $\gamma\gtrsim 10^5$ are propagating rectilinearly and
generally are not seen.  A quantitative analysis of the UHECR
flux in this scenario will be given elsewhere.

\section{Discussion and conclusions}

The nature of GRB engines is still unknown. There are
observational indications that they are astrophysical objects: about 25
GRBs are reliably found to be located in galaxies, probably in regions of
star formation; at least in one case GRB is identified with a supernova
(SN 1998 bw).
The most popular now are
astrophysical models, with binary neutron star
mergers \cite{NaPaPi}, failed supernova \cite{Woo}, hypernova \cite{Pacz}
and supranova \cite{ViSte} being the front runners (for a critical review
see \cite{Vie99}). All these models, however,
are not developed enough to give
quantitative predictions. They also share the difficulty with explaining
the large beaming factor required for GRBs.

In contrast, the cosmic string model presented here allows one to obtain
quantitative predictions for the main observational characteristics of
GRBs.  In this paper we developed a deliberately simplified model,
which is characterized by a single free parameter (the energy scale
$\eta$ of
symmetry breaking, or $\alpha=k_gG\eta^2$) and by three other physical
quantities,
relatively well restricted (the magnetic field in filaments
and sheets $B_0$, the epoch of galaxy formation $z_B$, and the
density of baryonic matter in the filaments and sheets, a quantity not
critical for the predictions). Nevertheless, the model correctly
accounts for the GRB rate, and for the range of GRB
fluences and durations. It may also explain the
short-time structure of GRBs, the diffuse $X$- and $\gamma$-ray
backgrounds, and the ultrahigh-energy cosmic rays.

The string model predicts recurrence of GRBs with a period of
$T_l\sim 50 \alpha_{-8}(1+z)^{-1/2}$ yrs. Very short bursts may
have much shorter recurrence periods, perhaps as short as a few
years.  Observation of these
repeaters is a challenge for the future detectors with a high efficiency
for detection of short bursts.

Another testable prediction of the model is that GRBs should be
accompanied by strong bursts of gravitational radiation with a
characteristic waveform.

It must be emphasized that our model involves a number of
simplifying assumptions.  All loops at cosmic time $t$ were assumed to
have the same length $l\sim\alpha t$ with $\alpha\sim k_g G\eta^2$,
while in reality there should be
a distribution $n(l,t)$.  The evolution law (\ref{B}) for $B(z)$ and
the assumption of $f_B={\rm const}$ are also oversimplified.  A more
realistic model should also account for a spatial variation of $B$.
Being basically a one-parameter model, our model may
predict spurious correlations between the GRB characteristics.
In particular, the $S\propto \tau_{GRB}^{-1}$ correlation, suggested by
Eq.(\ref{duration}), holds only at a fixed redshift and tends to be
washed out when the redshit distribution, the loop length distribution
$n(l,t)$, and the inhomogeneous
spatial distribution of the magnetic field are taken into account.

Our model meets basically one difficulty: it predicts too low
GRB rate from galaxies. This discrepancy could be explained if our
model strongly underestimates the capture rate of string loops by
galaxies.  For example, if $\alpha\gg k_g G\mu$, then the loops are
non-relativistic and may be effectively captured by galaxies.
Another possibility is that our model could
describe some subclass of the sources not associated with galaxies.
Such a subclass could include the short-duration GRBs for which host
galaxies are not found. In this case, the model needs a smaller $\alpha$, as
discussed in section IIIB. In contrast to the prediction of our model,
the short bursts do not show strong deviation from the Euclidean distribution.
This could be due to observational selection effects, since the faint
short-duration GRBs which form this subclass have a low  detection
efficiency in BATSE. Alternatively, it could be another subclass of
no-host GRBs.

On the other hand, GRBs from cusps have properties which distinctly
distinguish them from those produced by collapsars: they are
periodically repeating on the scale of a few decades for majority
of GRBs and on the scale of a few years for faint bursts
($S \lesssim 10^{-8}$~erg/cm$^2$) with short duration
$\tau_{\rm GRB} \lesssim 20$ ms. The next generation of GRB detectors
can examine this prediction.

\section*{Acknowledgement}
We are grateful to Roger Blandford, Ken Olum and Bohdan Paczynski for useful
discussions.  The work
of VB and BH was supported in part by INTAS through grant No 1065 and
the work of AV by the National Science Foundation (USA).


\begin{thebibliography}{99}
\bibitem{invlength} The actual length of the loop changes as the loop
oscillates.  The invariant length is defined as $l=E/\mu$, where $E$
is the loop's energy in its center-of-mass frame.
\end{thebibliography}
\end{document}